\documentclass[twocolumn,showpacs,preprintnumbers,amsmath,amssymb]{revtex4}
\usepackage[dvips]{graphicx}
\usepackage{amssymb,amsfonts,amsmath}

\begin{document}

\title{Statistical mechanics of the international trade network}

\author{Agata Fronczak, Piotr Fronczak}
\email{agatka@if.pw.edu.pl}
%\homepage[http://www.if.pw.edu.pl/$\sim$agatka]
%\thanks{}
\affiliation{Faculty of Physics, Warsaw University of Technology,Koszykowa 75, PL-00-662 Warsaw, Poland}

\date{\today}

\begin{abstract}
Analyzing real data on international trade covering the time interval 1950-2000, we show that in each year over the analyzed period the network is a typical representative of the ensemble of maximally random weighted networks, whose directed connections (bilateral trade volumes) are only characterized by the product of the trading countries' GDPs. It means that time evolution of this network may be considered as a continuous sequence of equilibrium states, i.e. quasi-static process. This, in turn, allows one to apply the linear response theory to make (and also verify) simple predictions about the network. In particular, we show that bilateral trade fulfills fluctuation-response theorem, which states that the average relative change in import (export) between two countries is a sum of relative changes in their GDPs. Yearly changes in trade volumes prove that the theorem is valid.
\end{abstract}


\maketitle

\section{Introduction}
In recent years, an extensive research effort has been devoted to analyzing the structure, function, and dynamics of the international trade network (ITN) \cite{2003_PRE_Serrano, 2003_PhysA_Li, 2004a_PRL_Garlaschelli, 2007_Econ_Serrano, 2008_JStatMech_Bhattacharya, 2009_PRE_Fagiolo, 2010_NewJPhys_Garas, 2010_JEvolEcon_Fagiolo} from a complex network perspective \cite{2009_Science_nets}. The knowledge of topological properties of this network and its evolution over time is not only important per se (e.g., because it enhances our descriptive knowledge of the stylized facts pertaining to the ITN), but it may also be relevant to a better explanation of macroeconomic dynamics \cite{2007_JIntBus_Kali, 2009_Science_Schweitzer, 2009_AdvCS_Schweitzer,2007_EPJB_Reichardt, 2008_AdvCS_Reyes, 2010_EconInq_Kali,2007_EPJB_Garlaschelli, 2009_ChinPL_Wen}.

Here, we use quantitative and numerical (data-driven) methods originating from statistical mechanics to describe and predict the behavior of ITN. We analyze a set of year-by-year trade relationships between all countries of the world, covering the time interval $1950-2000$. Although the total number of countries and the overall economic conditions influencing the network change over the course of the period, in each year ITN is shown to be a typical representative of the ensemble in which every network, $G$, is assigned the probability \cite{2004_PRE_Park, 2009_PRL_Garlaschelli} $P(G)\propto e^{-H(G)}$; where $H(G)=\sum_{i,j}\theta_{ij}w_{ij}$ plays the role of network Hamiltonian; $w_{ij}$ represents the volume of trade between two countries, $i$ and $j$; $\theta_{ij}\propto(x_ix_j)^{-1}$ is the field parameter conjugated to this trade connection; and $x_ix_j$ corresponds to the product of the GDPs of the trade partners.

Behind the descriptive power of our approach (which has been confirmed in a number of tests reported in this article consisting in comparison of GDP-driven Monte Carlo simulations of the trade network with real data on ITN), it is also important to stress the predictive abilities of the model. In particular, we show here that bilateral trade fulfills a simple fluctuation-response theorem \cite{2006a_PRE_Fronczak}.  Supported by the well-known qualitative findings about economic crises, we argue that the theorem provides valuable quantitative insights into dynamics of ITN in the time of crises, and, in the future, may be used to uncover mechanisms underlying the emergence of worldwide crises.

\section{Data description}

The results reported in this work are based on the trade data collected by K. S. Gleditsch \cite{dataset} that contains, for each world country in the period $1950-2000$, the detailed list of bilateral import and export volumes. The data are employed to build a sequence of matrices, $\mathbf{W}(t)$, corresponding to snapshots of weighted directed ITN in the consecutive years, $t=1950,\dots$. In the network, each country is represented by a node and the direction of links follows that of wealth flow. The entry, $w_{ij}(t)$, of the trade matrix, $\mathbf{W}(t)$, represents the weight of the directed connection. From the point of view of the country denoted by $i$, $w_{ij}(t)$ refers to the volume of export to $j$, while, from the point of view of the country labelled by $j$, it is seen as the volume of import from $i$. Precisely due to differences in reporting procedures between countries, when analyzing the data one often encounters small deviations between exports from $i$ to $j$ and imports from $i$ to $j$. To overcome the problem, in our analysis we define $w_{ij}(t)$ as the arithmetic average of the two values.

In this article, apart from trade matrices, which contain complete but often excessively detailed information about ITN, we also use several other quantities that make theoretical description of the network possible. In particular, to characterize the economic performance of a country we use its total GDP value, $x_i(t)$. To get the whole set of total GDPs, $\{x_i(t)\}$, we simply multiply the GDP \emph{per capita} by the population size of each country \cite{PWT}. Furthermore, to describe the intensity of the trade relationships of a country we define the so-called out-strength, $s_i^{out}(t)$, and in-strength, $s_i^{in}(t)$, of the corresponding node. The quantities are calculated as the total weight of connections (outgoing and incoming, respectively) that are attached to the node and they represent total volumes of export and import of the considered country in a given year, $t$.

All the data used in this study are given in millions of contemporary U.S. dollars. The disturbing effects of inflation are ruled out in a natural way by the fact that whenever the variables, $x_i(t)$ or $w_{ij}(t)$, are used in the calculations, they are intrinsically divided by the normalization constant that equals the sum of all variables of a given type, i.e. $X(t)=\sum_ix_i(t)$ or $T(t)=\sum_{i,j}w_{ij}(t)$, respectively. (In what follows, whenever there is no confusion we will often omit explicit time dependence of the quantities). As a byproduct, the above observation allows one to present the main results of this article in a very concise way, with the help of relative quantities defined as follows:
\begin{equation}\label{xiit}
\xi_i=\frac{x_i}{X},\;\;\;\sigma_i^{out}=\frac{s_i^{out}}{T},\;\;\; \sigma_i^{in}=\frac{s_i^{in}}{T},\;\;\; v_{ij}=\frac{w_{ij}}{T}.
\end{equation}

\section{Exponential random graphs: Structural Hamiltonian of the international trade network}

The first contributions dealing with international trade from a complex network perspective used a binary-network approach, in which one has assumed that a (possibly directed) link between any two countries is either present or not, depending on whether the trade volume that it carries is larger than a given threshold \cite{2003_PRE_Serrano, 2003_PhysA_Li, 2004a_PRL_Garlaschelli}. With reference to this line of research we would like to highlight the paper by Garlaschelli and Loffredo (2004), published in \emph{Physical Review Letters} \cite{2004a_PRL_Garlaschelli}. In the paper, the authors used the same real-world data to analyze an unweighted and undirected version of ITN, i.e. a network of partnership in trade. They have shown that the total GDP of a country, $x_i$, can be identified with the fitness variable \cite{2002_PRL_Caldarelli} that, once a form of the probability of trade connection between two countries is introduced, completely determines the expected structural properties of this network. This in turn implies that ITN viewed as a binary network is a typical representative of exponential random graphs \cite{2004_PRE_Park}. Furthermore, one can expect that the same holds true for the weighted version of this network \cite{2009_PRL_Garlaschelli, comment}.

To verify this conjecture, we start by considering the most general ensemble of directed weighted networks, which is described by the following Hamiltonian
\begin{equation}\label{Hw}
H(G)=\sum_i\sum_{j\neq i}\theta_{ij}w_{ij},
\end{equation}
with a separate parameter $\theta_{ij}$ coupling to each weighted connection. Our aim is to check whether the Hamiltonian is correct and, if so, how the parameters $\{\theta_{ij}\}$ depend on different indicators characterizing the global economy. To do this, we first examine the ensemble as it stands.

\begin{figure*}[ht]
\begin{center}
{\includegraphics[width=\textwidth]{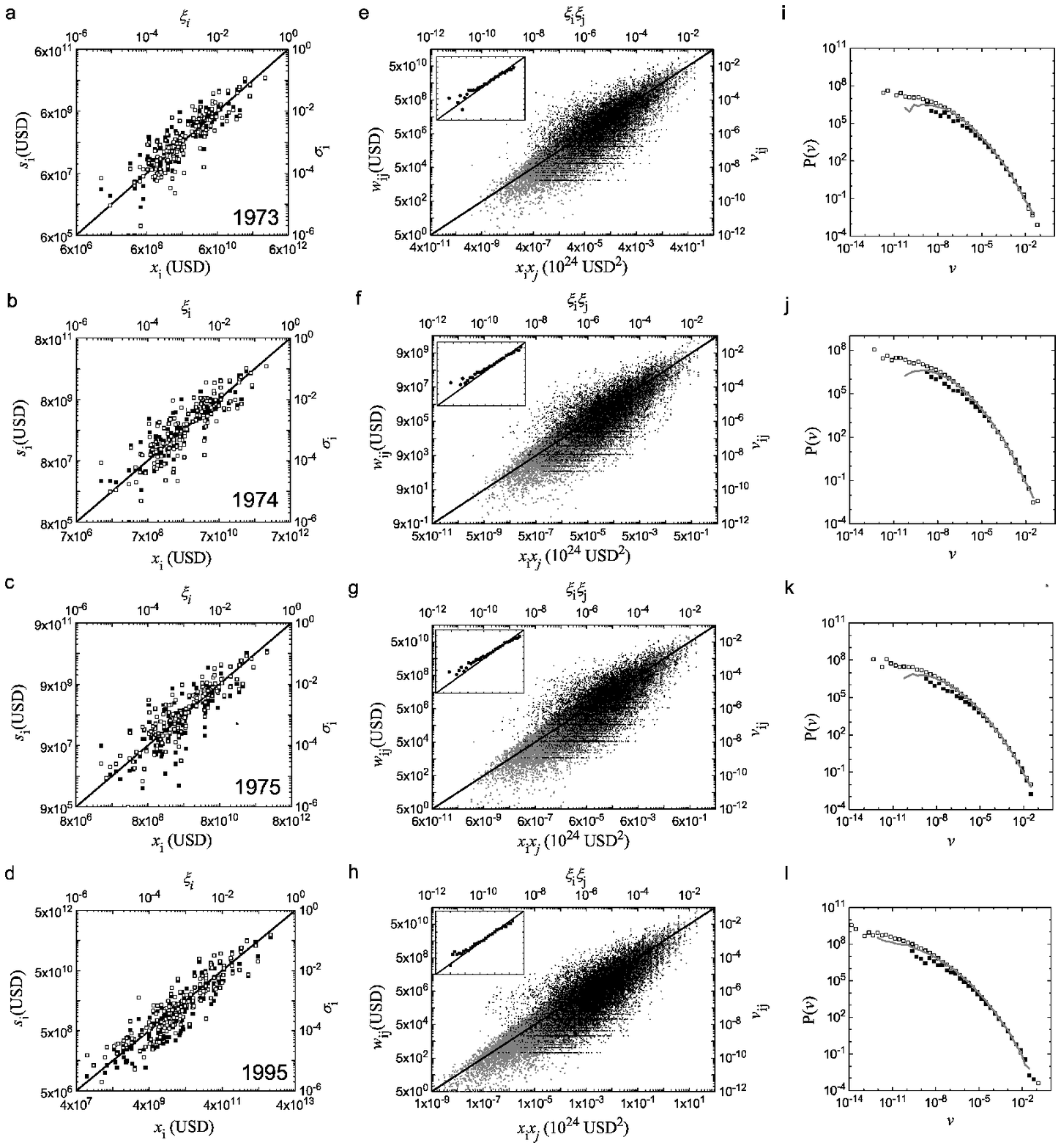}}
\caption{\textbf{Structural properties of single snapshots of ITN.} Diagrams placed in the same row refer to the same year. The data are shown in two ways, by using regular and relative quantities, cf.~Eq.~(\ref{xiit}). \textbf{a, b, c, d,} Total import and export volumes of all world countries in 1973, 1974, 1975, and 1995 vs. GDP (filled and open points correspond to imports and exports, respectively) and their comparison with the expected values described by Eqs.~(\ref{siITN}) and (\ref{sigmaITN}) (solid straight lines). \textbf{e, f, g, h,} Bilateral trade flows in the following years vs. the product of the trading countries' GDPs (points) as compared with their theoretical prediction based on Eqs.~(\ref{meanw}) and (\ref{omegaITN}) (line). Black points correspond to real data, while gray points represent trade volumes obtained from GDP-driven Monte Carlo simulations. In the insets of all the panels (e, f, g, and h), comparison of the expected theoretical values (straight lines) to mean values of the real data is shown. Since trade flows smaller than a given threshold are rarely specified in economic reports (in particular, the considered data set \cite{dataset} does not contain trade volumes smaller than 1000 USD), the cloud of black points cover smaller area than the one corresponding to numerical simulations. \textbf{i, j, k, l} Distributions of trade volumes in the considered years. Filled and open squares correspond to real and simulated data, respectively. The solid lines represent distributions of expected trade flows which, for each pair of countries in a given year, can be calculated using Eqs.~(\ref{meanw}) or (\ref{omegaITN}).}\label{fig1}
\end{center}
\end{figure*}

Thus, given that $w_{ij}$ is a real number greater than $0$ (as is true for trade volumes), the partition function of this ensemble can be written as
\begin{equation} \label{Zw}
Z(\{\theta_{ij}\})=\prod_i\prod_{j\neq i}\int_0^\infty e^{-\theta_{ij}w_{ij}} dw_{ij}=\prod_i\prod_{j\neq i}\frac{1}{\theta_{ij}}.
\end{equation}
This allows us to rewrite the probability of a network as $P(G)=e^{-H(G)}/Z=\prod_{i}\prod_{j\neq i}p_{ij}$, where
\begin{equation} \label{pijw}
p_{ij}(w_{ij})=e^{-\theta_{ij}w_{ij}}\theta_{ij}
\end{equation}
is the probability that there is a directed link of weight $w_{ij}$ from $i$ to $j$. The expression for $p_{ij}$  that we arrive at is the exponential distribution. Its mean value,
\begin{equation} \label{meanw}
\langle w_{ij}\rangle=1/\theta_{ij},
\end{equation}
can be used to calculate the average values of a node's out- and in-strength \begin{equation}\label{meansi}
\langle s_i^{out}\rangle=\sum_{j\neq i}\langle w_{ij}\rangle=\sum_{j\neq i}\frac{1}{\theta_{ij}} \;\;\;\mbox{and}\;\;\;\langle s_i^{in}\rangle=\sum_{j\neq i}\frac{1}{\theta_{ji}}.
\end{equation}

At this stage one can start to make comparisons of theoretical predictions with the empirical data on international trade. With good reason, it is convenient to begin by putting together Eq.~(\ref{meansi}) and the corresponding empirical relations (see Fig.~\ref{fig1}~a,~b,~c, and~d):
\begin{equation}\label{siITN}
\langle s_i^{out}\rangle=Ax_i\;\;\;\;\;\mbox{and}\;\;\;\;\;\langle s_i^{in}\rangle=Ax_i,
\end{equation}
where $A$ is the time-dependent parameter having the same value for both out- and in-strength of the nodes. Analyzing the expressions, one finds that the simplest way to merge the theoretical approach with real data is to assume a multiplicative form of the parameter $\theta_{ij}$, i.e.
\begin{equation}\label{thetaij}
\theta_{ij}=\theta_i\theta_j,
\end{equation}
where $\theta_i$ and $\theta_j$ represent some single-node characteristics controlling for the potential ability of the two nodes to be connected. One should note that the symmetric expression for $\theta_{ij}$, Eq.~(\ref{thetaij}), is consistent with observations made by other authors, showing the symmetric character of bilateral trade relations (see e.g.~\cite{2004a_PRL_Garlaschelli, 2010_JEvolEcon_Fagiolo}).

To calculate explicit values of all the parameters $\{\theta_i\}$, one just has to insert Eq.~(\ref{thetaij}) into the theoretical formula for $\langle s_i^{out}\rangle$, Eq.~(\ref{meansi}), and then equate the obtained relation to the empirical one, Eq.~(\ref{siITN}). (The analogous calculations can be done for $\langle s_i^{in}\rangle$.) As a result, one gets the following expression: $\langle s_i^{out}\rangle=\sum_j\theta_j^{-1}/\theta_i=Ax_i$, which, when summed over $i$, yields an important relation between theoretical and empirical quantities describing ITN: $T=\left(\sum_i\theta_i^{-1}\right)^2=AX$, from which it follows that
\begin{equation}\label{thetaij2}
\theta_{i}=\frac{1}{\sqrt{T}}\frac{X}{x_i}=\frac{1}{\sqrt{T}}\frac{1}{\xi_i}\;\;\;\;\; \mbox{and}\;\;\;\;\;\theta_{ij}=\frac{1}{T}\frac{1}{\xi_i\xi_j},
\end{equation}
where $\xi_i$ has been introduced earlier, in Eq.~(\ref{xiit}). Now, one can insert (\ref{thetaij2}) into (\ref{meanw}). As a result one gets a stochastic version of the gravity model of trade \cite{1979_AmEconRev_Anderson, 1985_Econ_Bergstrand, 2004_book_Feenstra},
\begin{equation}\label{gravity0}
\langle w_{ij}\rangle= \frac{T}{X^2}x_ix_j.
\end{equation}
In comparison to the standard gravity equation for bilateral trade volume between $i$ and $j$,
\begin{equation}\label{gravity1}
w_{ij}=const \frac{x_ix_j}{f(D_{ij})},
\end{equation}
where $f(D_{ij})$ represents transport costs in international trade, which depend on geographic distance between the two trading countries.

The expressions, Eqs.~(\ref{thetaij2}), together with other relative parameters defined in Eq.~(\ref{xiit}), can be used to rewrite the most important results of this approach in a very concise way. In particular, as described in terms of trade, the average out- and in-strength of a node, Eq.~(\ref{meansi}), when divided by the world's trade volume, $T$, turns out to be equal to the country's share in the world's GDP, $X$, i.e.
\begin{equation}\label{sigmaITN}
\langle\sigma_i^{out}\rangle=\langle\sigma_i^{in}\rangle=\xi_i.
\end{equation}
In a similar fashion, the average weight of a directed connection when divided by $T$, is given by
\begin{equation}\label{omegaITN}
\langle v_{ij}\rangle=\xi_i\xi_j=1/\eta_{ij}.
\end{equation}
Finally, the structural network Hamiltonian, Eq.~(\ref{Hw}), when written in relative variables has the following form
\begin{equation}\label{Hwr}
H(G)=\sum_{i}\sum_{j\neq i}\eta_{ij}v_{ij}.
\end{equation}

To verify the correctness of the assumed network Hamiltonian, a series of data-driven Monte Carlo simulations employing the Metropolis algorithm has been performed. The obtained results, for four years: 1973, 1974, 1975, and 1995, are shown in Figure~\ref{fig1} (diagrams placed in the same row refer to one year marked in the left chart). The first three years correspond to the 1973 oil crisis which caused high inflation and a global recession which affected all aspects of living within the 1970s. The figure shows that global economic crises, such as the one triggered by the 1973 oil crisis, do not affect the quality of our approach. In particular, in panels e, f, g, and h, in the Figure ~\ref{fig1} sets of all bilateral trade volumes recorded in a given year versus the product of the trading countries' GDPs is compared with the corresponding set of weights of directed connections in a typical network of the considered ensemble. Although the two sets (clouds) of points are quite dispersed, they overlap significantly, and their shape is well-described by Eq.~(\ref{omegaITN}). Moreover, as shown in panels i, j, k, and l in the same figure, the distributions of trade volumes within these clouds fit very well with each other and agree with the distribution of expected trade flows, $P(\langle v_{ij}\rangle)$, testifying in favor of our simple, yet realistic, approach.

\begin{figure*}[ht]
\begin{center}
{\includegraphics[width=\textwidth]{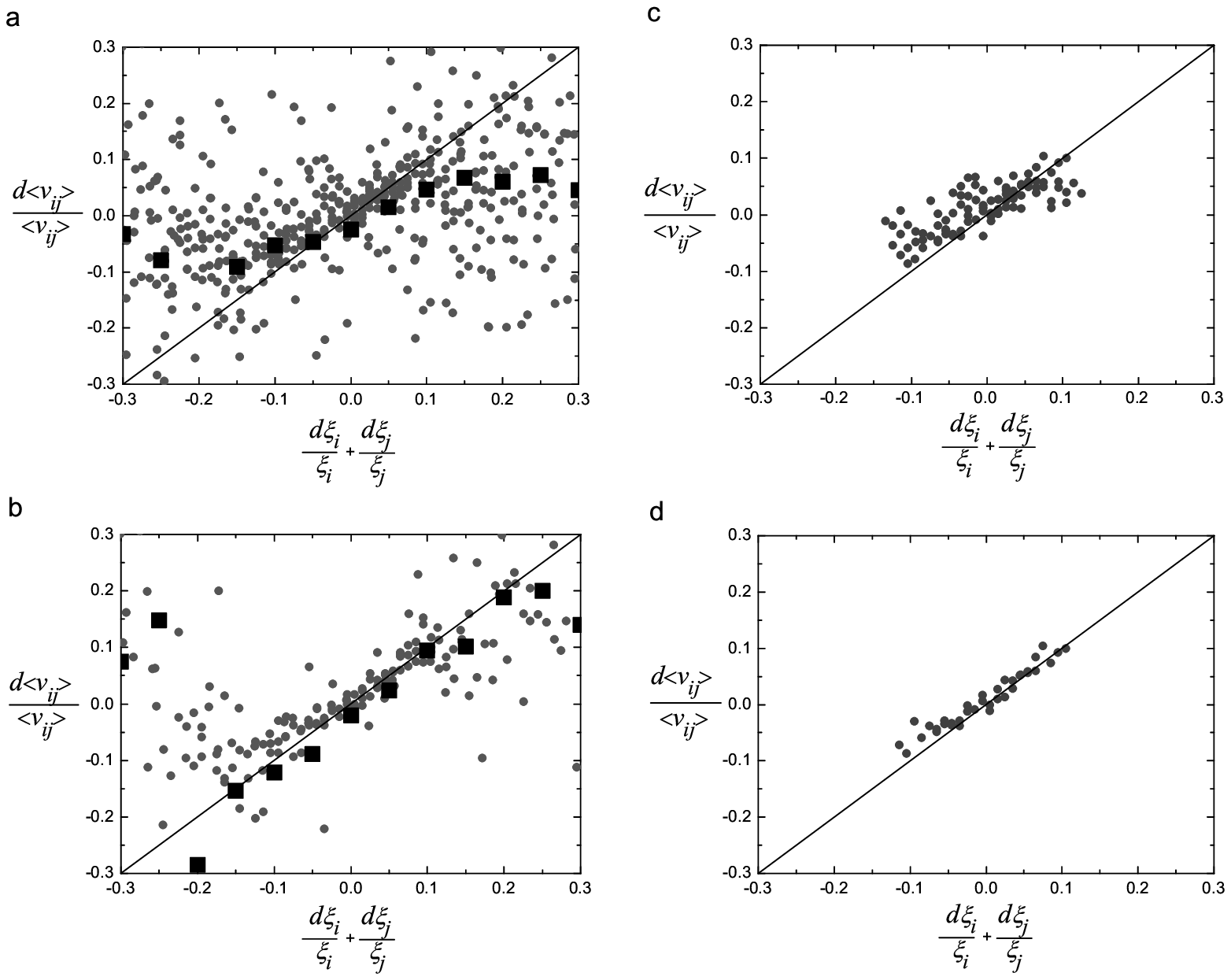}}
\caption{\textbf{Fluctuation-response theorem for ITN.} To confirm validity of the theorem, for each connection in the trade network at the turns of $1950-2000$, one has calculated quantities corresponding to both sides of Eq.~(\ref{frt3}): the relative change in normalized trade volume, $dv_{ij}/v_{ij}\simeq (v_{ij}(t+1)-v_{ij}(t))/v_{ij}(t)$, and the sum of relative changes in GDP of trading countries, $d\xi_i/\xi_i+d\xi_j/\xi_j$, where $d\xi_i/\xi_i\simeq (\xi_i(t+1)+\xi_i(t))/\xi(t)$. Then, the results were grouped according to similarities in both: local fields conjugated to the corresponding trade flows, $\eta_{ij}$, Eq.~(\ref{omegaITN}), and their year-by-year changes, $d\xi_i/\xi_i+d\xi_j/\xi_j$, characterizing the initial economic conditions influencing trade and the magnitude of the applied perturbation, respectively. Each group of connections, $V(m,n)$, was characterized by two integers, $m$ and $n$. The trade flow, $v_{ij}$, was classified as belonging to $V(m,n)$ if $m-1\leq\ln\eta_{ij}<m$ and $n-1\leq100\;d\eta_{ij}<n$. (Note that the grouping with respect to $m$ is in fact logarithmic binning with respect to the expected trade volume, Eq.~(\ref{omegaITN}), while the parameter $m$ describes linear binning with respect to the sum of percentage changes in relative GDPs of trade partners.) The gray circles shown in the figure correspond to geometric averaging of the results pertaining to single connections of ITN. The averaging was employed over the predefined groups, $V(m,n)$. The black squares correspond to grey points averaged over the horizontal axis. More precisely, the upper left panel \textbf{a,} shows all the data such as they are described above. The upper right panel \textbf{b,} presents the same data with the additional condition: it illustrates the fluctuation-response theorem as applied to trade volumes whose expected share, $\langle v_{ij}\rangle$,  Eq.~(\ref{omegaITN}), in the global trade was greater than $10^{-4}$. The lower left panel, \textbf{c,} presents those points from the panel a, for which the number of single trade volumes contributing to the geometric average was greater than $1000$. Finally, in the lower right panel \textbf{d,} the gray circles which meet the additional conditions specified in panels b, and c, are shown.}\label{fig2}
\end{center}
\end{figure*}

\section{Susceptibility of trade volume to changes in GDP: Fluctuation-response relation for ITN}

Extensive comparisons between real data on international trade and its GDP-driven Monte Carlo simulations show that, although the total number of world's countries, $N(t)$, and their GDPs, $\{x_i(t)\}$, change over the analyzed period of $50$ years, ITN is continuously well-characterized by the same Hamiltonian. This means that the time evolution of this network may be considered as a continuous sequence of equilibrium states (i.e. quasi-static process) that is yearly sampled by the national reporting procedures. Furthermore, since differences between snapshots of ITN in the consecutive years are rather small, one can expect that they could be described with the help of liner response theory, of which the simplest (but not yet trivial) result is the fluctuation-response theorem \cite{2006a_PRE_Fronczak}.

In the case of exponential random graphs with the Hamiltonian given by Eq.~(\ref{Hwr}) the fluctuation-response theorem has the following form
\begin{equation}\label{frt1}
\langle v_{ij}^2\rangle-\langle v_{ij}\rangle^2=-d\langle v_{ij}\rangle/d\eta_{ij}.
\end{equation}
The l.h.s. of this expression describes fluctuations in relative weight, $v_{ij}=w_{ij}/T$, of the directed connection from $i$ to $j$, whereas its r.h.s. characterizes susceptibility of this link to its conjugated local field, $\eta_{ij}$. The susceptibility is defined as the derivative of $\langle v_{ij}\rangle$ with respect to $\eta_{ij}$ and describes what happens with the expected trade volume, $\langle v_{ij}\rangle$, when one changes the parameter $\eta_{ij}$, which determines external conditions for the bilateral exchange.

Taking into account the exponential distribution of weights, $p(v_{ij})=e^{-\eta_{ij}v_{ij}}\eta_{ij}$, which follows from Eq.~(\ref{pijw}) and whose variance equals the square of the mean, the fluctuation-response theorem, Eq.~(\ref{frt1}), can be transformed into the formula
\begin{equation}\label{frt3}
\frac{d\langle v_{ij}\rangle}{\langle v_{ij}\rangle}=-\frac{d\eta_{ij}}{\eta_{ij}}=\frac{d\xi_i}{\xi_i}+\frac{d\xi_j}{\xi_j}.
\end{equation}
Written in such a way, the theorem states that relative changes in normalized (i.e. divided by $T$) bilateral trade volumes can be estimated on the basis of changes in the GDP of trade partners. Yearly changes in import/export volumes between different countries prove that the fluctuation-response theorem for ITN is correct (see Figure~\ref{fig2}). Although, we have obtained Eq.~(\ref{frt3}) as the fluctuation-response relation for exponential random graphs, it is obvious that it can be also obtained as the logarithmic derivative of Eq.~(\ref{omegaITN}).

Relying on Eq.~(\ref{frt3}) one can, for example, expect that a decline of, say, $2$ percent in the relative GDP of a country, given that its trade partners do not change their share of the world's GDP, will translate into a similar decline in all its bilateral trade volumes. The example shows that the theorem can be used to make simple predictions about the world-wide diffusion of trade-based economic perturbations. Furthermore, assuming that the structure of ITN is a proxy for meaningful financial linkages between countries, the expression may also help one to understand how financial ripples originating in one country propagate to other countries, giving rise (or not) to global financial contagion. In particular, in the light of this theorem, it becomes apparent that a crisis is amplified if the epicenter country is better integrated into the trade network \cite{2010_EconInq_Kali}. This happens because the decline of its GDP, through the proportional downward effect on trade and financial linkages, affects more countries. It also becomes clear that the impact of a crisis on any target country is cushioned if the country in question is better integrated into ITN, due to the fact that, in the case of such countries, the decline of only one bilateral trade volume does not significantly influence GDP.

\section{Concluding remarks}

\emph{The current economic crisis illustrates a critical need for new and fundamental
understandings of the structure and dynamics of economic networks.} This sentence opens the perspective article from the special issue of \emph{Science} \cite{2009_Science_nets} on complex systems and networks entitled "Economic networks: The new challenges" by F.~Schweitzer et al. In their article \cite{2009_Science_Schweitzer}, the authors summarize what we know and what we need to know about different economic networks (including ITN) to reduce the risk of global depression and to design effective strategies to promote economic recovery \cite{2009_NP_Lux}. Our approach to ITN is in line with this challenging research area. Having the mathematically tractable yet realistic model of ITN introduced here, and given its quasi-static time evolution, we believe that, apart from the fluctuation-response theorem, Eq.~(\ref{frt3}), other well-known results of non-equilibrium statistical physics \cite{2004_PT_Ruelle} may be applied to estimate recession (or economic growth) impact on international trade.

The exponential random graphs are in common use within the statistics and social network analysis communities as a practical tool for modeling networks for a few decades. Although such modeling gives a good qualitative description of the data, to our best knowledge, there is no evidence so far that with such approach one could achieve excellent quantitative accuracy and reproducibility of real world phenomena.

\begin{acknowledgments}
This work was supported by the Polish Ministry of Science, Grant No. 496/N-COST/2009/0.
\end{acknowledgments}

\end{document}